\documentstyle[psfig,aps,prl]{revtex}
\begin{document}

\title{
Rolling up of graphite sheet: Energetics of shell formation
}
\author{Slava V. Rotkin, Robert A. Suris}
\address{
Ioffe Physico--Technical Institute
of Russian Academy of Sciences,
\\ 26 Politekhnicheskaya St., St.Petersburg 194021, Russia}
\date{15 August 1999} \maketitle

\begin{abstract}
The energetics of transformation of a planar fragment of a
graphite monolayer into a spherical cluster is studied.
The path considered is that a flat cluster rolls up
into a segment of a spherical shell. The energy
landscape
of the process is presented. A simple model, formerly invented
for calculating the carbon nanocluster formation energy, is used to
evaluate the energies of intermediate states.
Although the spherical-shell closed cluster has the lowest energy, curving
of a plane fragment into a segment has an energy
barrier. The barrier height goes to zero for clusters
with the number of atoms greater than some $N_{\rm th}$, for
which the cluster size is found analytically.
\end{abstract}


The aim of the work is to study the energetics of formation of
curved carbon nanoscale clusters (CNC).
 If the synthesis conditions
more or less correspond to equilibrium, the
energy considerations give an insight into the mechanisms
of cluster formation,
which explains the persistent
theoretical interest in the CNC
energetics\cite{toman,ters,ssc,luc}.
This led us to propose a new phenomenological model
%
%
 in Refs.\cite{reno,mex}. In the frame of the model, only three
parameters allow estimating the formation energies for a
variety of clusters within unified analytical approach.  When
these parameters are furnished by quantum--chemical
calculations or
extracted from experiment, a number of general laws related
to the cluster stability is deduced.

In the paper, we consider
clusters which can serve as intermediates for
rolling of a flat fragment of a graphite monolayer
into a closed sphere. Tube formation will be discussed
elsewhere\cite{nano99,jcm99}

The continual approximation is supposed to be valid
for a curved graphite-like surface.
The model assumes that the
carbon bonds are the same for all atoms in any cluster
(excepting bonds belonging to a pentagonal defect of the
pristine honeycomb lattice of 2D graphite).
We argue that, at least
for the large cluster\cite{cont},
the continual approach works well.
We extrapolate the results of the continual approach and apply
the model to an interesting case of comparatively small clusters
with the number of atoms, $N$, of about a hundred. We
state that, despite the simplification, the model
adequately estimates the
CNC formation energy for a tube,
sphere, and capsule \cite{mex,bost}.

The paper proceeds as follows: Section \ref{sec:curv} deals with
the fundamentals of our model, illustrated by examples of
a carbon nanotube and spherical cluster
(a detailed description of our approach was presented in
Refs.\cite{reno,mol,diss}).
In Section
\ref{sec:sph} we discuss the energetics of a possible
mechanism by which a spheroidal cluster can be formed by
rolling--up of a piece of a graphite-like plane. Finally, a
summary is given.

\section{
Energy of curved graphite surface}
\label{sec:curv}

{\bf Three
phenomenological parameters}
were used \cite{mol,diss} to calculate the
additional energy of CNC formation, as compared with the
known specific
energy of an infinite graphite sheet (graphene).
The energy of any carbon cluster with a curved surface
is decomposed into additive terms.
Each term is given by
some characteristic energy (the model parameter) multiplied by a
number depending solely on the cluster geometry.
This essential
simplification of the calculation is based on natural
reasoning.  There are 3 sources bringing at least 3 parameters
into the energy (excepting the "zero energy" of
an atom belonging to
an infinite graphite sheet; this term
is excluded from further expessions).
The first additional term comes from
dangling bonds. The corresponding energy parameter,
$E_{b}$, is the dangling
bond energy.
An extra energy due to a change of the degree of a
hybridisation of an electron orbit
on the curved surface is described using
the second parameter, $E_{c}$. The last parameter, $E_{5}$, is
the total energy of twelve pentagonal rings. This parameter
reflects that an electron belonging to the pentagon must have
energy different from
that in a graphite-like hexagon.  Hence, it
partly takes into
account non-equivalent bonds. Note that this is
a topologically determined donation to the total energy of any
closed cluster \cite{3a}.

In general, the curvature of a surface is a function
of a point. The curvature is constant and equal to
$3/R^2$ for a sphere of radius $R$, \
and $9/8R^2$ for a tube of the same radius.
For the sake of clarity, we
consider here only
the surface with constant curvature.
       One easily obtains
the formation energy of an infinite carbon
tube. The specific energy of dangling bonds is
negligible since it goes to zero with the tube size going
to infinity. A
tube has no pentagon, because a cylinder is topologically
equivalent to a plane.
Hence, the 
energy has only one term,
the
"curvature energy"
$E=E_c \; 9N/8 R^2$,
where $N$ is the number of atoms
and $R$ is the dimensionless tube
radius.
Henceforth, all lengths will be measured in bond lengths
($b\simeq 1.4 $\AA~ not supposed to vary according to
the model).  We compared the tube energy with the
relevant results of
other computations \cite{ssc}. This gave the first
parameter of the model: $E_{c}\simeq 0.9$~eV~(see also
\cite{reno,mol}).

Most likely, the energy of a dangling bond in
CNC does not differ too much
from the value for graphite, and
therefore, we took $E_{b}=2.36$~eV.
The total
energy associated with the dangling bonds is proportional to
the cluster perimeter length (length of the open
boundary of the CNC lattice) multiplied by
the density of dangling bonds along
the perimeter, which in turn depends on the local
geometry. One can imagine, for example, a finite tube as
a parallelogram, carved from hexagonal graphene, which has its
opposite sides glued \cite{ssc,drs}.
%
%
 This parallelogram
has a variable number of sites on the open perimeter $\zeta$
(dangling bond density),
depending on its orientation relative to the lattice vector. None
the less the density
varies in a narrow region. If we take
the geometrical factor in $\zeta$ to be unity for the "zigzag"
tube, then it becomes $2/\sqrt{3}$ for the "armchair" tube; one
can easily calculate $\zeta$ for the bond density of a
specific tube lying between these extreme examples.

It was shown
in our previous papers \cite{reno,mol,diss,mrs98},
that there occurs {\bf competition }
in the formation energy of a finite tube
between two terms with $E_{c}$ and
$E_{b}$ (the energy associated with dangling bonds can not be
neglected for a tube of finite length).
The tube energy reads as: $E(N,R)=E_c \; 9N/8R^2+E_b \;
4\pi\zeta R/\sqrt{3}$.
It is advisable to make the
total perimeter $4\pi R$ shorter
in order to
diminish the number of dangling bonds. The radius decrease costs
an extra total energy owing to stronger
curvature. We called the
cluster having the minimal energy at a certain number of atoms
"an optimal cluster".
The optimal cluster is the configuration
governed by energy consideration
at any fixed $N$. (The nanotube energetics and the
tube formation road from a flat graphene were discussed in
details in Ref.\cite{jcm99}).
       Thus, we constructed a phenomenological
approach and predicted the optimal cluster shape, based  on the
energetics of any nanotube without performing a
quantum-chemical calculation each time.

The third parameter of the model was determined by
fitting the formation
energy of a spherical cluster to the
experimental value for $C_{60}$. The total energy
of a sphere consists of two
components \cite{mol}.
The first constant component is the energy of 12 pentagons.
The sphere energy also
includes
the total curvature energy which is independent of the
radius in this case (because the number of atoms being
proportional to squared curvature, $1/R^2$). The second component
is a correction excluding pentagonal bonds from the total
count of curved bonds. Therefore, the energy is as follows:
\begin{equation}
E_{\rm sph}=\left( E_{5}+\frac{16 \pi
E_{c}}{\sqrt{3}}\right) -\frac{N_s}{N}E_{c}
\label{22}
\end{equation}
where we introduce a characteristic number
$N_s=2\times60\times16\pi/3\sqrt{3}\simeq 1161$.
The number $N_s$
includes the number of bonds belonging to the
pentagonal rings in the fullerene (cf. also \cite{mol})
and the (dimensionless) curvature of the surface times
the (dimensionless) area of the sphere covered by hexagons.
We took the experimental value of C$_{60}$ formation energy
from Ref.\cite{exp} and fitted our last parameter
$E_{5}\simeq17.7$~eV, which takes into account all 12
pentagons of the spheroidal closed cluster.
The sphere energy evidently remains
positive
(for the given model
parameters $E_c$=0.9 eV and $E_5$)
for the smallest
spherical cluster considered here, $C_{60}$.
 Then, comparing the experimental formation energy of $C_{70}$
with our
calculation, we find that the difference is about 1\%
and all three parameters are
self--consistent.

\section{\bf How sphere is rolled up}
\label{sec:sph}

A sphere has the minimal curvature energy between closed
nanoclusters of any definite number of atoms, $N=$const. It seems
interesting to compare the energies of
{\bf closed clusters} with the energy
of {\bf an open fragment} having
dangling bonds. For an open CNC, the
shorter the perimeter (number of dangling bonds), the stronger
the curvature. The decreasing energy of dangling bonds
makes favourable a small rolled--up cluster with a shorter
perimeter. The curvature energy demands that the CNC be
extended and flat. As a result of the energy competition, the
system decreases its total energy via elimination of
dangling bonds owing to the fact that the infinitely
large sphere is a configuration resulting in the global minimum
of the total CNC energy \cite{mol}.

We consider below
a sphere (of area $S_o$) with a round hole of
angular size $\Omega$ varying from 0 to
$4\pi$, from the sphere to the round piece of the plane
(see Fig.1). This means that we choose a flat CNC with
{\bf the minimal perimeter} (round) and
a closed CNC with {\bf the minimal curvature}
(sphere) at fixed area or number of atoms.
We name an intermediate cluster "an open
sphere".  The curvature energy decreases with the open
sphere surface area as $(1-\Omega/4\pi) S_o$.
This can be seen from the
expression for the number of atoms in an ideal
graphene sphere with
a round hole:
$N=(1-\Omega/4\pi)~16\pi~R^2/3\sqrt{3} $.
It is also natural to feather out
the dependence of the topological
energy $E_5(\Omega)$ in the hole size $\Omega$ (actually
varying stepwise, each step correspondes to
creating a pentagonal defect) and to substitute
a linear dependence. Instead of choosing a
specific way to create a defect and to place it
in the cluster lattice,
the simplest uniform distribution
of pentagons over the open sphere surface is used here
\cite{unpubl}. This is nearly equivalent to placing pentagons
as far apart
as possible. The kinetics of the process \cite{unp2} is
not touched on in this paper, as well as the pentagon--pentagon
distance optimised elsewhere \cite{nano99}.

In terms of the independent variables $R$ and
$\Omega$ (or $N$ and $\Omega$ alternatively)
the energy of the open sphere reads as:
\begin{eqnarray}
\left.
\begin{array}{c}
E(N,\Omega) =   \displaystyle
\left( 1-\frac{\Omega}{4\pi}\right)
\left( E_{5}+E_{c} \left[ \frac{16 \pi}{\sqrt{3}}
-\frac{120}{R^2} \right] \right)
+\displaystyle
\frac{4\pi R E_{b}}{\sqrt{3}\zeta}\sqrt{\frac{\Omega}{4\pi}}
\sqrt{1-\frac{\Omega}{4\pi}} = \\
\displaystyle =\left( 1-\frac{\Omega}{4\pi}\right)
\left[ E_{5}+\frac{16 \pi E_{c}}{\sqrt{3}}
-E_{c}\frac{N_s}{N}     \left( 1-\frac{\Omega}{4\pi}
\right)\right]
+ \displaystyle
2\left[ E_{5}+\frac{16 \pi E_{c}}{\sqrt{3}}\right]
\frac{1}{\sqrt{N_{\rm th}}}
\sqrt{\frac{N\Omega}{4\pi}}
\end{array}
\right. \label{open}
\end{eqnarray}
where $\zeta$ is a geometrical multiplier of about unity
\cite{corr}, and
we introduce $N_{\rm th}$, a "threshold" size of planar
cluster, given by:
$\displaystyle
\frac{\sqrt{\pi\sqrt{3}}}{\zeta} \frac{E_{b}}
{2[E_{5}+\frac{16\pi}{\sqrt{3}}E_{c}]}
=\sqrt{N_{\rm th}^{-1}}$. Its meaning
is discussed just below.
The first term in Eq.(\ref{open}) is the
energy of an incomplete sphere. The second
is the dangling
bond energy.
Of course, the accuracy of the model is poorer for small $N$
and strong curvature \cite{small}.
As it is seen from Figure 2, the energy of
the open sphere with hole size of neither 0 nor
$4\pi$ has a maximum at fixed number of atoms when $N\le
N_{\rm th}\simeq 254$.
Partial differentiation of
Eq.(\ref{open}) with respect to $\Omega$ (at fixed $N$) gives
an analytical expression for the size of the open sphere with the
maximal energy.  The function is given
(with high accuracy for $N>100$) by:
\begin{equation}
\Omega_B \simeq 4\pi \displaystyle \left( 1-
\left(1-\frac{N}{N_{\rm th}}\right)
\frac{N}{\displaystyle
N+4N_s/
\left[\frac{E_{5}}{E_{c}}+\frac{16\pi}{\sqrt{3}}\right] }
\right).
\label{line}
\end{equation}
The line given by Eq.(\ref{line})
intersects the abscissa axis at $\Omega=4\pi$ and $N_{\rm th}$,
named "the threshold flat cluster" in the following
sense.

Let us consider a set of CNCs with $N<N_{\rm th}$ and
$4\pi<\Omega<0$: the energy of a cluster with $N= \rm const$
grows in moving away from both extremities of this interval and
has a
maximum at $\Omega=\Omega_B$. That is,
going from $\Omega=4\pi$ to $\Omega=0$ (where the
minimum of the energy occurs at fixed $N$), we pass through a
barrier of height $\delta E(N)= E(N,\Omega_B)-E(N,4\pi)$ 
decreasing with $N$ and tending to zero at $N_{\rm th}$.
Therefore, as follows from energy consideration solely,
planar clusters with $N\ge N_{\rm th}$
can be rolled up without any
barrier. Below $N_{\rm th}$, the flat CNC is
metastable to the rolling--up.
Suppose we have initially
some distribution of planar graphite fragments
$f(N)$
(which is consistent with the two--step--model
of carbon cluster
formation proposed in \cite{eb} where
nucleation and growth processes were separated).
It is
energetically favourable to increase the number of atoms,
but this
process depends on availability
of the building material (carbon atom concentration).
One can suppose that in the second step of
synthesis the only conformation is possible
between different clusters
(even through some exchange of carbons).

We
consider below that the change in number of atoms is
suppressed.  Suppose that the most energetically favourable
process is rolling--up of a flat CNC into a sphere \cite{roads}.  
Then
clusters with $N>N_{\rm th}$ are rolled--up
for energetical reason.  At some finite temperature
$T$, the probability to roll up a smaller cluster is
exponentially small: $W\propto \exp(-\delta E/T)$,
where $\delta E$ is the barrier height
(shown in Figure 2) depending on $N$ and $\Omega$. This is the
difference between the energy of the given
flat cluster and that of
a "barrier" cluster with the same number of atoms.

As a result of rolling--up, the initial size distribution of
planar clusters is changed. We discussed above only
a single path of
conformation (rolling--up into the spherical-shell
cluster).  For this reason we can only speculate about the
distribution of the rest of planar clusters. It has a maximum
depending on temperature and characteristic number of
atoms in the initial distribution $f(N)$, being
determined by the
conditions of two distinct synthesis regimes and,
therefore, treated as independent .




In summary,
we proposed a heuristic
model for calculating the energy of
a carbon nanocluster formation. Within the model,
we use only three phenomenological
parameters extracted from computer
simulations and experimental data. The model allows
evaluating the formation energies of various CNCs and finding
the preferable cluster shape.

We compared the formation energies of the most
preferable closed cluster (spherical cluster) and the most
preferable flat fragment (round piece of graphene). We
conclude that for energy reasons rolling--up of a flat
fragment is always favourable, since the
energy of a closed cluster
is the lowest. We predict that a small enough flat
graphite cluster can be metastable with respect to rolling into
a shell.  The energy barrier for rolling is calculated
and shown to disappear for the clusters containing more than
three hundred atoms.

{\bf Acknowledgments. }
This work was partially supported by RFBR grants no.
96--15--96348 and 98--02--18117.

The supposed path of spherical-shell closed
cluster formation: from a flat round fragment of graphene,
$\Omega=4\pi$ (right), via open spheres with $0<\Omega<4\pi$
(center), to a shell, $\Omega=0$ (left). The number of atoms is
kept constant. Therefore, the linear size decreases.

~\\~


~\\~

The energy landscape of the process of
rolling--up. The barrier height
for rolling--up is shown. The
energy of a flat round fragment of graphene is lower than that
of an intermediate state, the open sphere,
depending on hole size $\Omega$ and number of
atoms $N$. The barrier disappears at $N>N_{\rm th}$, as described
in the text.


\eject

~\vskip 1cm
\begin{figure}[h]
\centerline{ \psfig{figure=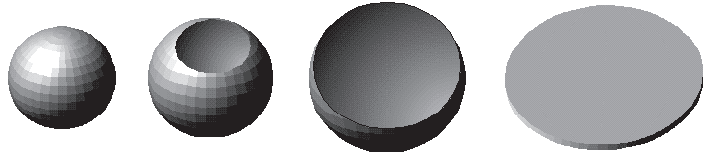,width=10cm}}
~\vskip 1cm
\caption{\label{fig:path}
The supposed path of spherical-shell closed
cluster formation: from a flat round fragment of graphene,
$\Omega=4\pi$ (right), via open spheres with $0<\Omega<4\pi$
(center), to a shell, $\Omega=0$ (left). The number of atoms is
kept constant. Therefore, the linear size decreases.}
\end{figure}

\begin{figure}[h]
\centerline{ \psfig{figure=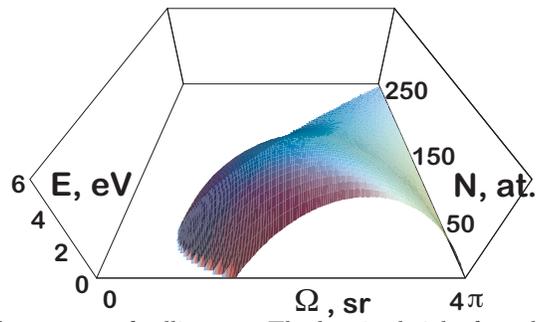,width=7cm}}
\caption{\label{fig:barrier}
The energy landscape of the process of
rolling--up. The barrier height
for rolling--up is shown. The
energy of a flat round fragment of graphene is lower than that
of an intermediate state, the open sphere,
depending on hole size $\Omega$ and number of
atoms $N$. The barrier disappears at $N>N_{\rm th}$, as described
in the text. } \end{figure}


\begin{references}
\bibitem{3a}
The topological Gauss-Bonnet theorem allows creating a closed
polyhedron with 3-coordinated vertices from any number of
hexagons and exactly 12 pentagons. Therefore, the energy
parameter $E_5$ is named a topological one.
\bibitem{unpubl}
The position of pentagonal defects in the energetically
optimal cluster lattice is not known exactly. The
Isolated-Pentagon-Rule works for the closed
clusters. Even if we suppose uniform distribution to be valid
for an open cluster, a
possible reconstruction of the open
perimeter
is to be considered.
This may lead to additional tension of surface
and
reshaping of the cluster through pentagon migration.
Also, the kinetics of pentagon
diffusion seems to be non-trivial \cite{unp2}. These effects are
far beyond the scope of the paper and will be discussed
elsewhere.
\bibitem{corr}
Of course, one can not make a round cut from the hexagonal
lattice of graphene. It will have hexagonal angles on each
"tooth". The factor $\zeta$ reflects the actual
geometry, being  1 for a perfectly smooth perimeter
with fixed minimal density of dangling bonds $1/\sqrt{3}b$,
equal to the "zigzag" tube density.
\bibitem{small}
If one extrapolates Eqs.(\ref{open}) and (\ref{line}) to
very small clusters ($N<60$), at least Eq.(\ref{22}) fails
owing to the fact that
all atoms belong to the pentagons and some of the bonds
become inequivalent.
The strain energy has to increase with increasing $1/R^2$.
Thus, the model underestimates the energy of
smaller clusters. However, one can use the model for
small flat and slightly curved
fragments of plane
with some pentagons.  That is,
the lower the pentagon density (the number of pentagons per
CNC area, see \cite{cont} above),
the better the model accuracy.
\bibitem{roads}
Indeed, the formation of tubes seems to be a
competitive process. It was shown \cite{luc}
to have a large energy barrier and therefore will be 
suppressed comparing with non--threshold
rolling--up in a spherical shell.
\end{references}
\end{document}